# Capacitance between Two Points on an Infinite Grid

**J. H. Asad**[1]**, R. S. Hijjawi**[2]**, A. Sakaji**[3]**, and J. M. khalifeh**[1,4]

[1] Dep. of Physics, University of Jordan, Amman-11942-Jordan.
E-mail: jhasad1@yahoo.com.
[2] Dep. of Physics, Mutah University, Jordan.
E-mail: hijjawi@mutah.edu.jo.
[3] Dep. of Basic Sciences, Ajman University, UAE
E-mail: info@ajss.net.

[4] Corresponding author EMAIL: jhasad1@yahoo.com.

## Abstract

The capacitance between two adjacent nodes on an infinite square grid of identical capacitors can easily be found by superposition, and the solution is found by exploiting the symmetry of the grid. The mathematical problem presented in this work involves the solution of an infinite set of linear, inhomogeneous difference equations which are solved by the method of separation of variables.

## 1. Introduction

The electric circuit networks has been studied well by Kirchhoff's[1] more than 150years ago, and the electric-circuit theory is discussed in detailed by Van der Pol and Bremmer[2] where they derived the resistance between any two arbitrary lattice sites. Later on, at the ends of the eighties of the last century the problem is revived by Zemanian[3], where he investigated the resistance between two arbitrary points in an infinite triangle and hexagonal lattice networks of identical resistor using Fourier series. For hexagonal networks, he discovered a new method of calculating the resistance directly from the networks.

The problem is studied again by many authors[4–11]. For example, Cserti[8] and Cserti et. al[9] introduced in their papers how to calculate the resistance between arbitrary nodes for different lattices where they presented numerical results for their calculations. In recent works, we used Cserti's method to calculate theoretically the resistance between arbitrary sites in an infinite square and Simple Cubic (SC) lattices and experimental comparison with the calculated values are presented [10,11]. Finally, Wu[12] studied the resistance of a finite resistor network where the resistance between two arbitrary nodes is obtained in terms of the eigenvalues and eigenfunctions of the Laplacian matrix associated with the finite network.

Little attention has been paid to infinite networks consisting of identical capacitances *C*. Van Enk[13] studied the behavior of the impedance of a standard ladder network of capacitors and inductors where he analyzed it as a function of the size of the network. In this paper we investigated analytically and numerically the capacitance between arbitrary lattice sites in an infinite square grid using the superposition principle. Also, the asymptotic behavior is studied for large separation between the two sites. An investigation of infinite complicated lattices and of lattices with missing capacitor (bond) is in progress.

The physical situation is illustrated in Fig. 1. An infinite number of identical capacitors of capacitance *C* are connected to form an infinite square grid. The problem is to find the capacitance between arbitrarily spaced nodes. The basic approach used here is similar to that used by Paul[14].

Let a charge $Q$ enter the grid at a node $r_o$ and let it comes out of the grid at a distant point. Removing the return point to infinity then the problem is invariant under $90^o$ rotation, so the charge flowing through each of the four capacitors connected to the node will be equal. Therefore each one of them will carry a charge of $\frac{Q}{4}$. Thus, the resulting voltage drop between node $r_o$ and an adjacent node $r$ will be $\frac{Q}{4C}$.

Now, consider the case where a charge $Q$ entering the grid at a distant point and exiting at the adjacent node, $r$. Again, the charge flowing will be $\frac{Q}{4}$, and the voltage drop from $r_o$ to $r$ will be given by $\frac{Q}{4C}$.

The superposition of the above two problems results in a new problem where a charge $Q$ entering the node $r_o$ and exiting the adjacent node $r$ with a net voltage drop equal to $\frac{Q}{2C}$. The distant point is eliminated since the net charge there is zero, therefore the capacitance between adjacent nodes is $2C$.

## 2. Node Voltages Analysis

Consider an infinite square network consisting of identical capacitors such as that shown Fig. 1. Let the nodes be numbered from minus infinity to plus infinity in each direction, and let the voltage at the node *(m,n)* be denoted by $V_{m,n}$ . Applying Kirchhoff's laws at node *(m,n)*. Thus, one may write:

$$Q_{m,n} = (V_{m,n} - V_{m,n+1})C + (V_{m,n} - V_{m,n-1})C + (V_{m,n} - V_{m+1,n})C + (V_{m,n} - V_{m-1,n})C \,. \qquad (1)$$

If at node *(m,n)* the charge equal to zero, then Eq. (1) reduces to:

$$4V_{m,n} = V_{m,n+1} + V_{m,n-1} + V_{m+1,n} + V_{m-1,n} \,. \qquad (2)$$

Now, let a charge $Q$ enter the node *(0,0)* and leave at infinity. Then

$Q_{m,n} = 0$ Unless m= 0 and n= 0, $Q_{0,0} = Q$ . (3)

Or, we may write Eq. (3) as:

$$Q_{m,n} = \begin{cases} \text{Q}, & \text{m=n=0} \\ \text{zero}, & \text{otherwise.} \end{cases} \qquad (4)$$

Equations (1 and 2) are the finite difference equivalent of Poisson's and Laplace's equations, respectively.

## 3. Solution by Separation of Variables

Although the method of separation of variables is commonly applied to partial differential equations having suitable boundary conditions, it is equally applicable to difference equations[15].

Consider

$$V_{m,n} = \exp(m\alpha + in\beta) . \tag{5}$$

Substituting Eq. (5) into Eq. (2), we obtained the following:

$$4\exp(m\alpha + in\beta) = \exp((m+1)\alpha + in\beta) + \exp((m-1)\alpha + in\beta) +$$

$$\exp(m\alpha + i(n+1)\beta) + \exp(m\alpha + i(n-1)\beta). \tag{6}$$

The above equation can be rewritten as:

$$4\exp(m\alpha + in\beta) = \exp(m\alpha + in\beta)(\exp(\alpha) + \exp(-\alpha) + \exp(i\beta) + \exp(-i\beta)) . \tag{7}$$

Thus, Eq. (2) is satisfied provided that:

$$Cosh\alpha + Cos\beta = 2 . \tag{8}$$

Noting that our aim is to solve the problem with a source at *(*0,0*)*, this implies that:

$$V_{m,n} = V_{n,m} = V_{-m,n} = V_{m,-n} = V -_{m,-n} . \tag{9}$$

Therefore, take

$$V_{m,n} = \exp(-|m|\alpha)Cosn\beta + \exp(-|n|\alpha)Cosm\beta . \tag{10}$$

The above functions $V_{m,n}$ do not satisfy the source- free difference relation given by Eq. (2) along the lines *n=0 and m=0*. This is due to the absolute value sign in the exponential terms, so there will be residual charges entering or leaving the grid at each node along these lines.

To find the external charges $Q_{m,n}$ which produce the voltage pattern $V_{m,n}$, we may write from Eq. (1):

$$\frac{Q_{o,o}}{C} = 4V_{o,o} - V_{o,1} - V_{o,-1} - V_{1,o} - V_{-1,o};$$

$$= 8 - 4(Cos\beta + \exp(-\alpha)) . \tag{11}$$

Using Eq. (8), then Eq. (11) can be simplified as:

$$\frac{Q_{o,o}}{C} = 4Sinh\alpha\,. \tag{12}$$

In a similar way, for $n \neq 0$

$$\frac{Q_{o,n}}{C} = 4V_{o,n} - V_{o,n+1} - V_{o,n-1} - V_{1,n} - V_{-1,n}\,;$$

$$= 2Cosn\beta(2 - Cos\beta - Cosh\alpha) + 2\exp(|n|\alpha)\{2 - Cos\beta - Cosh\alpha\}\,. \tag{13}$$

Again, using Eq. (8) we can simplify Eq. (13) as:

$$\frac{Q_{o,n}}{C} = 2Cosn\beta Sinh\alpha\,. \tag{14}$$

## 4. Charge Entering at *(*0,0*)*

The assumed node voltage pattern $V_{m,n}(\beta)$ requires external charges $Q_{m,n}$ not only at node (0,0) but at all nodes for which either *m= 0 or n= 0*. Thus, it is necessary to form a superposition of such voltages with different values of $\beta$, to suppress all external charges except the one at (0,0*)*.

Now, let $V_{m,n}(\beta)$ having the following form:

$$V_{m,n} = \int_{o}^{\pi} F(\beta)V_{m,n}(\beta)d\beta\,. \tag{15}$$

where the limits of the integral have been chosen to cover the entire applicable range of values of $\beta$. The function $F(\beta)$ is an amplitude function that must be chosen to make $Q_{o,o} = Q$ and $Q_{o,n} = 0$, when $n \neq 0$.

Substituting Eq. (15) into Eq. (1), we obtained:

$$\frac{Q_{m,n}}{C} = \int_{0}^{\pi} F(\beta)Q_{m,n}(\beta)d\beta\,. \tag{16}$$

From Eqs. (12 and 13), we have:

$$\frac{Q_{o,o}}{C} = \int_{0}^{\pi} [F(\beta)4Sinh\alpha]d\beta\,. \tag{17}$$

and

$$\frac{Q_{o,n}}{C} = \int_{0}^{\pi} [F(\beta)2Sinh\alpha Cosn\beta]d\beta\,. \tag{18}$$

where $Q_{o,n} = Q_{o,-n} = Q_{n,o} = Q_{-n,o}$ .(i.e. $Q_{o,n}$ is symmetric).
The expression for $F(\beta)$ can be obtained by inspection as:

$$F(\beta) = \frac{Q}{C(4\pi Sinh\alpha)} . \tag{19}$$

which satisfies the above condition. Thus, Eq. (15) becomes:

$$V_{m,n} = \frac{Q}{4\pi C}\int_0^{\pi} \frac{V_{m,n}(\beta)}{Sinh\alpha} d\beta ;$$

$$= \frac{Q}{4\pi C}\int_0^{\pi} \frac{(\exp(-|m|\alpha) Cosn\beta + \exp(-|n|) Cosm\beta)}{Sinh\alpha} d\beta . \tag{20}$$

Note, that $\alpha$ is a function of $\beta$. They are related by Eq. (8).

## 5. Capacitance between Two Points in a Large Grid

We mentioned earlier that the capacitance between (0,0) and *(m,n)* could be obtained directly from the solution of the problem in which the charge *Q* enters at (0,0) and leaves at infinity. In terms of the node voltages, the capacitance $C_{m,n}$ can be written as:

$$C_{m,n} = \frac{Q}{2(V_{o,o} - V_{m,n})} . \tag{21}$$

Using Eq. (20), one may write Eq. (21) as:

$$C_{m,n} = \frac{2\pi C}{\int_0^{\pi} \frac{(2 - \exp(-|m|\alpha) Cosn\beta - \exp(-|n|\alpha) Cosm\beta)}{Sinh\alpha} d\beta} . \tag{22}$$

Or, one may write $C_{m,n}$ (see Appendix A) as:

$$C_{m,n} = \frac{\pi C}{\int_0^{\pi} \frac{(1 - \exp(-|n|\alpha) Cosm\beta)}{Sinh\alpha} d\beta} . \tag{23}$$

The integrals in Eqs. (22 and 23) have to be evaluated numerically. Results of $C_{m,n}$ for values of $(m,n)$ ranging from (0,0) – (10,10) are presented in Table 1 below, and here are some special cases:

(a) $C_{o,o}$

Using eq. (22) with *n=m=0*. Then

$$C_{o,o}=\frac{2\pi C}{\int_0^{\pi}\frac{(2-\exp(-0)Cos0-\exp(-0)Cosm0)}{Sinh\alpha}d\beta}=\frac{2\pi C}{0}=\infty\,. \qquad (24)$$

as expected. This can be explained as a parallel capacitance with zero separation between its plates.

(b) $C_{o,1}$

Again, using Eq. (22) with *n=0, m=1*. Thus

$$C_{o,1}=\frac{2\pi C}{\int_0^{\pi}\frac{(2-Cos\beta-\exp(-\alpha))}{Sinh\alpha}d\beta}=2C\,. \qquad (25)$$

where we have used $2-Cos\beta-\exp(-\alpha)=Sinh\alpha$.
The above result is the same as mentioned at the end of the introduction. We may explain this result as two identical capacitors connected together in parallel.

(c) Asymptotic form for large *m* or *n*

When either *m* or *n* are large, the exponential terms given in Eq. (23) become negligible except when $\alpha$ is very small. When $\alpha$ is very small;

$$Cos\beta=2-Cosh\alpha\approx1-\frac{\alpha^2}{2}\,. \qquad (26)$$

So that

$$\alpha\approx Sinh\alpha-\beta\,. \qquad (27)$$

Suppose *m* is large, then Eq. (23) can be rewritten as:

$$\frac{C_{m,n}}{C}=\pi[\frac{1}{\{\int_0^{\pi}\frac{(1-\exp(-|m|\beta)Cosn\beta)}{\beta}}+\frac{1}{\int_0^{\pi}\frac{(1/Sih\alpha-1)}{\beta}}+$$

$$\frac{1}{\int_0^{\pi}\frac{\exp(-|m|\beta)Cosn\beta}{\beta}}-\frac{1}{\int_0^{\pi}\frac{\exp(-|m|\beta)Cosn\beta}{Sinh\alpha}\}d\beta}]\,. \qquad (28a)$$

$$=\pi(\frac{1}{I_2}+\frac{1}{I_2}+\frac{1}{I_3})\,. \qquad (28b)$$

The first integral can be expressed in terms of the exponential integral Ein(z)[16]

$$Ein(z) = \int_0^z \frac{(1-\exp(-t))}{t} dt;$$

$$= \int_0^z \frac{(1-\exp(-\beta z/\pi))}{\beta} d\beta. \tag{29}$$

So, $I_1 = \mathrm{Re}\{Ein[\pi(n+im)]\}$.

The second integral can be integrated numerically, and it is found to be $I_2 = -0.1049545$. In the third integral the exponentials are negligible except for small values of $\alpha$ and $\beta$, and for those values $\alpha \approx Sinh\alpha \approx \beta$. So, $I_3$ can be neglected. Thus, Eq. (28) becomes:

$$\frac{C_{m,n}}{C} = \frac{\pi}{\mathrm{Re}\{Ein[\pi(n+im)]\}} - \frac{\pi}{0.1049545}. \tag{30}$$

For large values of its argument, $Ein(z) \to \ln z + 0.57721$. Therefore, Eq. (30) can be rewritten as:

$$\frac{C_{m,n}}{C} = \frac{\pi}{2\ln(n^2+m^2) + 4\ln\pi + 0.22038855}. \tag{31}$$

For reasonable values of *m* and *n*, the asymptotic form (i.e.: Eq. (31)) gives an excellent approximation, and from this equation we can show that as any of *m* and *n* goes to infinity then $\frac{C_{m,n}}{C} \to 0$, which can be explained as a parallel capacitance with infinite separation between its plates. Finally, it is clear from Eq. (23) that $C_{m,n} = C_{-m,-n}$ which is expected due to the inversion symmetry of the infinite square grid.

**6. Results and Discussion**

In this work, the capacitance between the site (0,0) and the site (m,n) ; in an infinite square grid consisting of identical capacitors is calculated using the superposition of charge distribution. The capacitance $C_{m,n}$ is expressed in an integral form which can be evaluated numerically or analytically.

The asymptotic form for the capacitance as m or/and n goes to infinity is investigated where it is shown that it goes to zero.

In Figs. (2-5) the capacitance is plotted against the site (m,n). Fig. (2) shows a three dimensional plot of the capacitance as a function of m and n. One can see from the figure that as m or/and n increases then $C_{m,n}$ decreases up to zero at infinity as expected before (i.e. see Eq. (31)).

Figures (3-5) show the capacitance $C_{m,n}$ as a function of the site (m,n) along the directions [10], [01] and [11]. From these figures we can see that the capacitance is symmetric along these directions, and this is due to the inversion symmetry of the infinite square grid. Also, the figures show how the capacitance $C_{m,n}$ goes to zero as any of m or n goes to infinity.

# Figure Captions

**Fig. 1 An infinite number of identical capacitors of capacitance *C* are connected to form an infinite square grid.**

**Fig. 2 The capacitance $C_{m,n}$ in terms of m and n for an infinite square grid.**

**Fig. 3 The capacitance $C_{m,n}$ in terms of the site along [10] direction.**

**Fig. 4 The capacitance $C_{m,n}$ in terms of the site along [01] direction.**

**Fig. 5 The capacitance $C_{m,n}$ in terms of the site along [11] direction.**

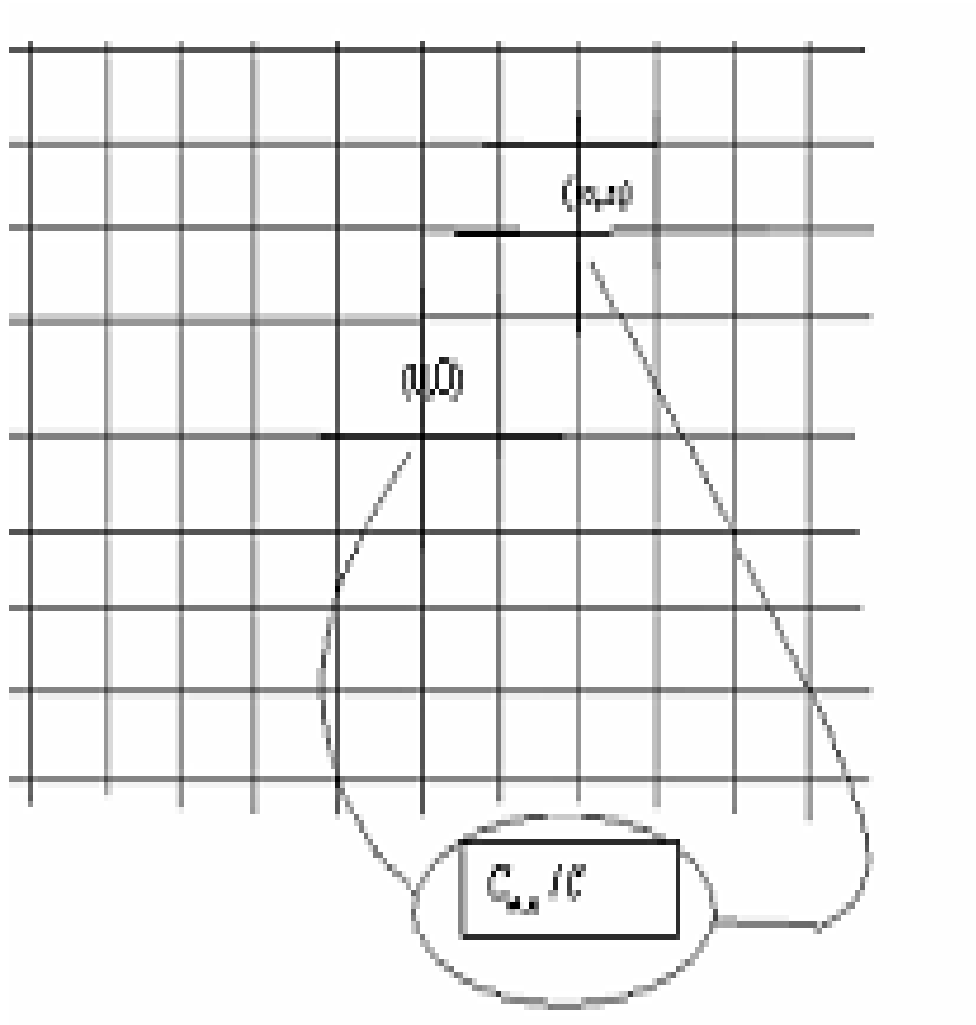

**Fig. 1**

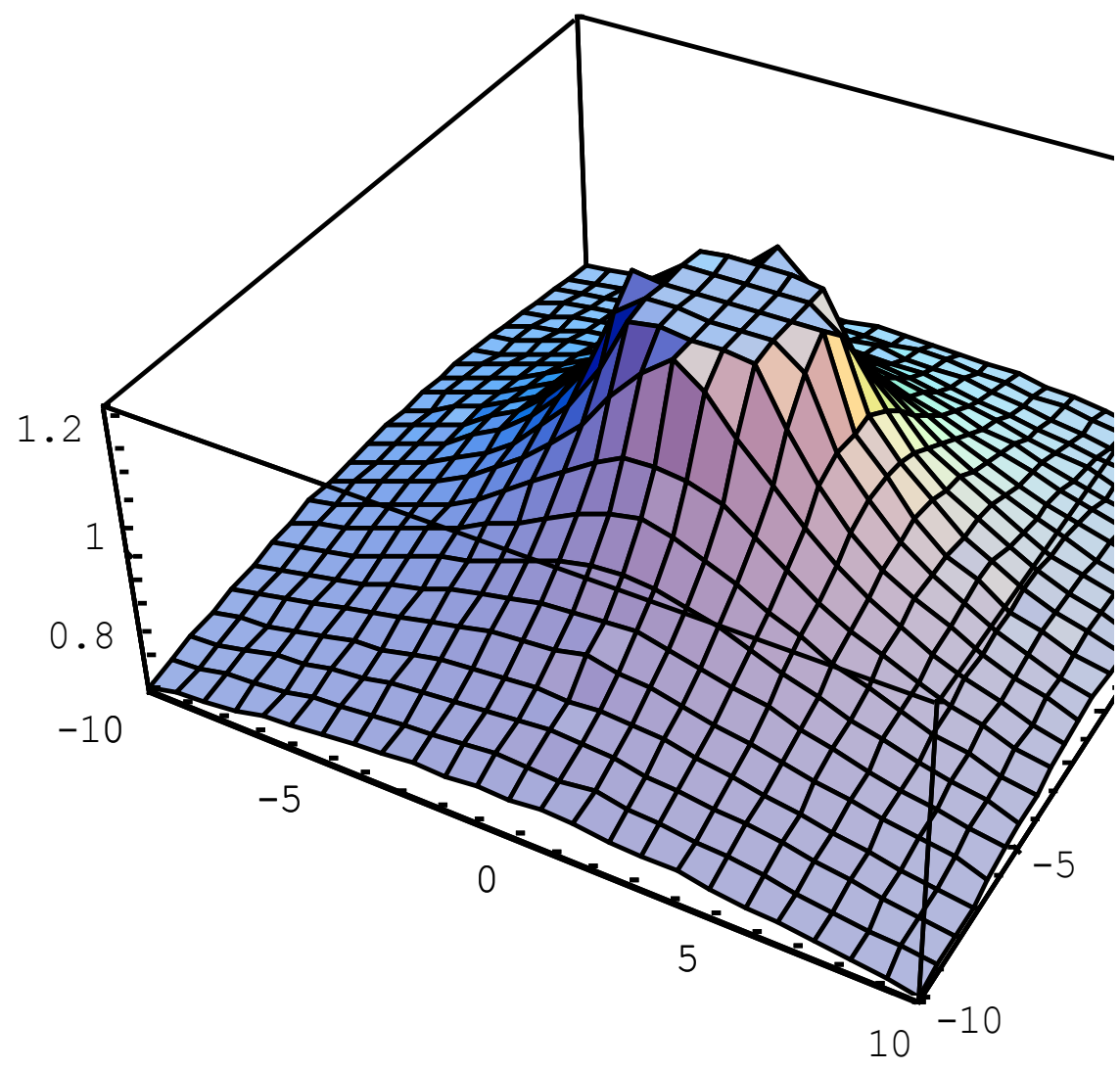

C(m,n)/C

**Fig. 2**

/C $C_{m,n}$

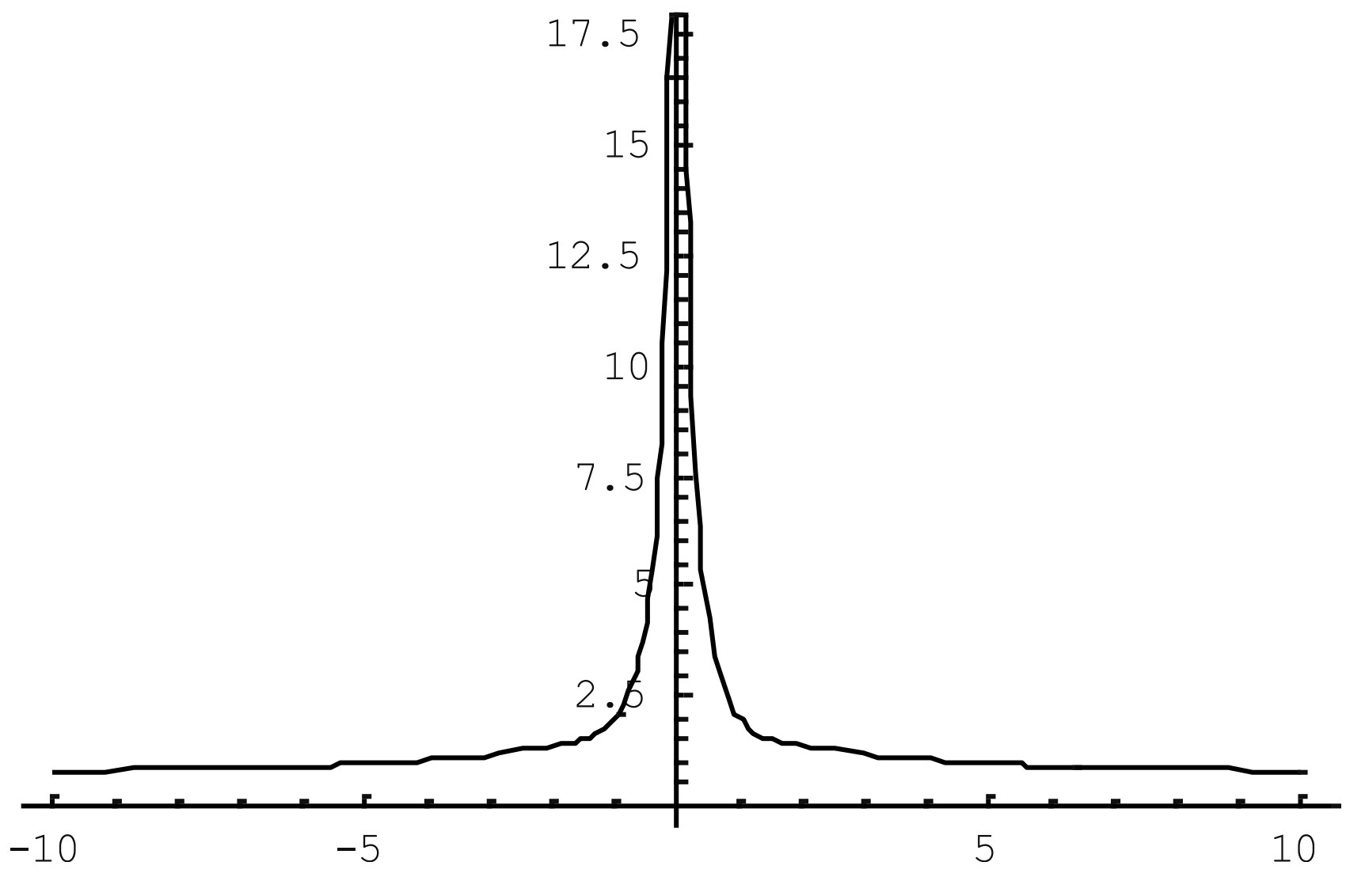

The site (m,0)

**Fig. 3**

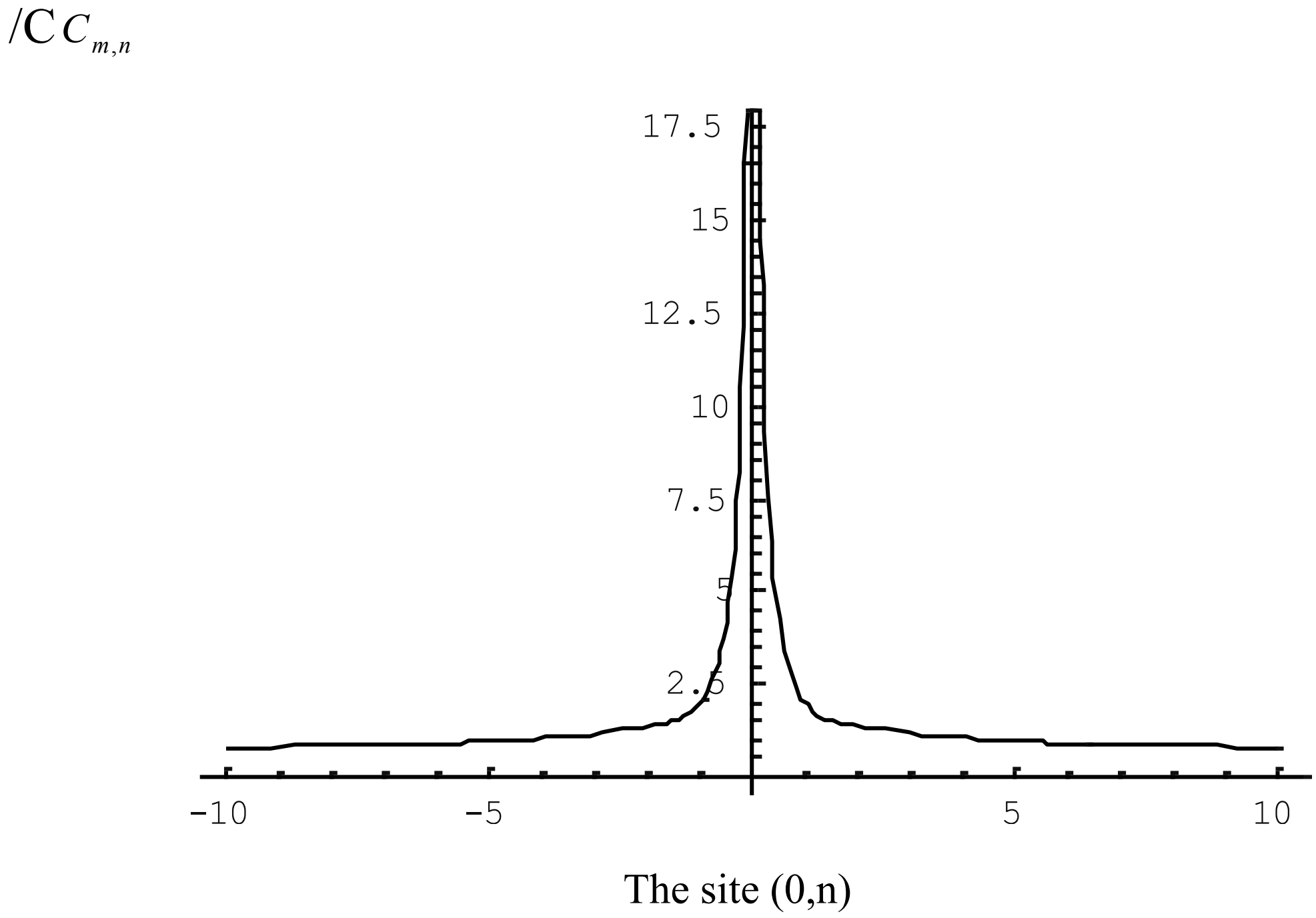
/C $C_{m,n}$
17.5
15
12.5
10
7.5
5
2.5
-10
-5
5
10
The site (0,n)

**Fig. 4**

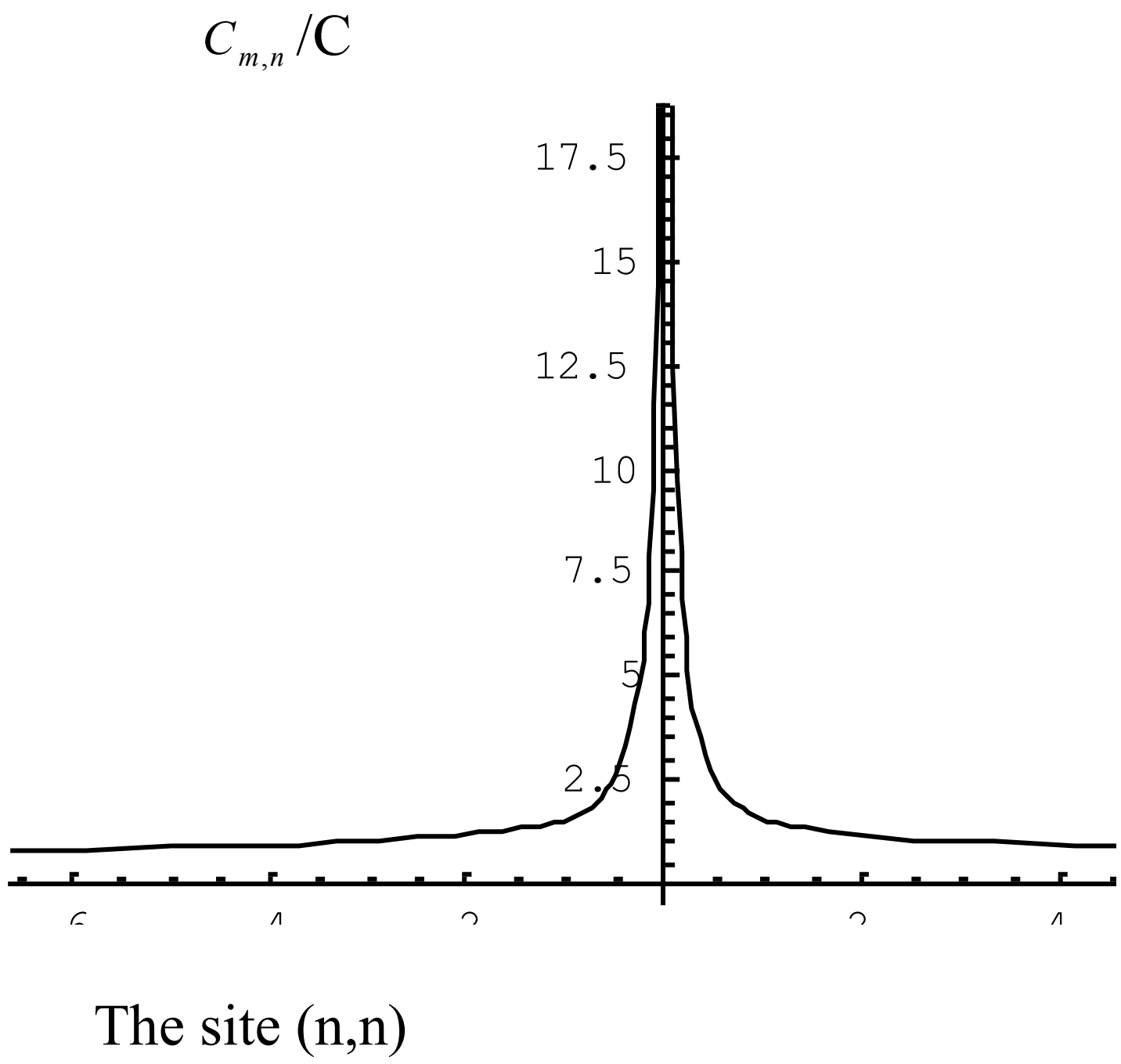
$C_{m,n}$ /C
17.5
15
12.5
10
7.5
5
2.5
The site (n,n)

**Fig. 5**

# Table Captions

**Table 1: Numerical values of $C_{m,n}$ in units of C for an infinite square grid.**

**Table 1**

| $C_{m,n}$ /C | (m,n) | $C_{m,n}$ /C | (m,n) | $C_{m,n}$ /C | (m,n) |
|---|---|---|---|---|---|
| 0.822545 | (9,1) | 0.907753 | (6,2) | ∞ | (0,0) |
| 0.818628 | (9,2) | 0.892285 | (6,3) | 2 | (1,0) |
| 0.812497 | (9,3) | 0.874193 | (6,4) | 1.5708 | (1,1) |
| 0.804631 | (9,4) | 0.85517 | (6,5) | 1.37597 | (2,0) |
| 0.795539 | (9,5) | 0.836326 | (6,6) | 1.29326 | (2,1) |
| 0.785687 | (9,6) | 0.882207 | (7,0) | 1.1781 | (2,2) |
| 0.775459 | (9,7) | 0.879628 | (7,1) | 1.16203 | (3,0) |
| 0.765148 | (9,8) | 0.872324 | (7,2) | 1.13539 | (3,1) |
| 0.754964 | (9,9) | 0.861357 | (7,3) | 1.08177 | (3,2) |
| 0.801699 | (10,0) | 0.847985 | (7,4) | 1.02443 | (3,3) |
| 0.800666 | (10,1) | 0.833344 | (7,5) | 1.04823 | (4,0) |
| 0.797649 | (10,2) | 0.818295 | (7,6) | 1.03649 | (4,1) |
| 0.792868 | (10,3) | 0.803421 | (7,7) | 1.00814 | (4,2) |
| 0.786636 | (10,4) | 0.850222 | (8,0) | 0.972869 | (4,3) |
| 0.779303 | (10,5) | 0.848397 | (8,1) | 0.937123 | (4,4) |
| 0.771206 | (10,6) | 0.843152 | (8,2) | 0.974844 | (5,0) |
| 0.762645 | (10,7) | 0.835079 | (8,3) | 0.968523 | (5,1) |
| 0.753862 | (10,8) | 0.824942 | (8,4) | 0.951831 | (5,2) |
| 0.745047 | (10,9) | 0.813496 | (8,5) | 0.929041 | (5,3) |
| 0.736338 | (10,10) | 0.801381 | (8,6) | 0.90391 | (5,4) |
| | | 0.789079 | (8,7) | 0.878865 | (5,5) |
| | | 0.776929 | (8,8) | 0.922313 | (6,0) |
| | | 0.823894 | (9,0) | 0.918443 | (6,1) |

## Appendix A

Instead of using the functions $V_{m,n}$ defined in Eq. (10), let us use the following functions

$$W_{m,n} = \exp(-|m|\alpha)Cos\beta . \quad \text{A1}$$

Equation (A1) is a source free everywhere except a long the line m=0. Thus, the external charge $Q_{0.n}$ can be written as:

$$\frac{Q_{o,n}}{C} = 2Cosn\beta Sinh\alpha . \quad \text{A2}$$

Now, let $V_{m,n}$ be given as:

$$V_{m,n} = \int_{o}^{\pi} F(\beta)W_{m,n}(\beta)d\beta . \quad \text{A3}$$

The corresponding expression for the external charge is

$$\frac{Q_{0,n}}{C} = \int_{0}^{\pi} F(\beta)2Sinh\alpha Cosn\beta d\beta . \quad \text{A4}$$

Using Fourier cosines series, one can write (using Eq. (A4))

$$2\pi F(\beta)Sinh\alpha = \frac{Q_{0,0} + 2\sum_{n} Q_{0,n}Cosn\beta}{C} . \quad \text{A5}$$

For our case considered here

$$2\pi F(\beta)Sinh\alpha = \frac{Q}{C} . \quad \text{A6}$$

Or, we may write

$$F(\beta) = \frac{Q}{2\pi CSinh\alpha} . \quad \text{A7}$$

Substituting Eqs. (A1 and A7) into Eq. (A3), we obtained:

$$V_{m,n} = \frac{Q}{2\pi C}\int_{0}^{\pi}\frac{\exp(-|m|\alpha)Cosn\beta}{Sinh\alpha}d\beta \, . \qquad \text{A8}$$

Finally, the capacitance $C_{m,n}$ can be obtained by inserting Eq. (A8) into Eq. (21). Thus we get:

$$\frac{C_{m,n}}{C} = \frac{\pi C}{\int_{0}^{\pi}\frac{1-\exp(-|m|\alpha)Cosn\beta}{Sinh\alpha}d\beta} \, . \qquad \text{A9}$$